\documentclass[10pt,fleqn]{article}

\usepackage{amsmath,amssymb}
\usepackage{graphicx}
\usepackage[top=0.75in, bottom=0.75in, left=0.75in, right=0.75in, dvips]{geometry}

\begin{document}

\title{Improvement of Renormalization-Scale Uncertainties Within Empirical Determinations of the $b$-Quark Mass}
\author{M.R.\ Ahmady,\thanks{Department of Physics, Mount Allison University, Sackville, New Brunswick, E4L 1E6, Canada}
~V.\ Elias,\thanks{Department of Applied Mathematics, The University of Western Ontario, London, Ontario  N6A 5B7, Canada}
~A.\ Squires,\thanks{Newman Laboratory of Nuclear Physics, Cornell University, Ithaca, NY  14853, USA}
~T.G.\ Steele,\thanks{Department of Physics and Engineering Physics, University of
Saskatchewan, Saskatoon, SK, S7N 5E2, Canada}
~Ailin Zhang\thanks{Department of Physics, Suzhou University, Suzhou, Jiangsu, 215006, China}
}

\maketitle
\begin{abstract}
Accurate determinations of the $\overline{\rm MS}$ $b$-quark mass $m_b\left(m_b\right)$ from
$\sigma\left(e^+e^-\to{\rm hadrons}\right)$ experimental data currently contain three comparable sources of uncertainty; the
experimental uncertainty from moments of this cross-section, the uncertainty associated with $\alpha_s\left(M_z\right)$, and
the theoretical uncertainty associated with the renormalization scale.  Through resummation of all logarithmic terms explicitly
determined in the perturbative series by the renormalization-group (RG) equation, it is shown that the renormalization-scale dependence
is virtually eliminated as a source of theoretical uncertainty in  $m_b\left(m_b\right)$.
This resummation also reduces the estimated effect of higher-loop perturbative contributions, further reducing the theoretical
uncertainties in $m_b\left(m_b\right)$. Furthermore, such resummation techniques
improve the agreement between the values of the $\overline{\rm MS}$ $b$-quark mass extracted from the
various moments of $R(s)=\sigma\left(e^+e^-\to{\rm hadrons}\right)/\sigma_{pt}$
[$\sigma_{pt}=4\pi\alpha^2/(3s)$], obviating the need to choose an
optimum moment for determining $m_b\left(m_b\right)$.
Based on this analysis, the resulting value of the $b$-mass is $m_b\left(m_b\right)=4.207\,{\rm GeV}\pm 40\,{\rm MeV}$, where the
dominant uncertainty now arises from the experimental moments. Resummation techniques are also shown to reduce renormalization-scale
dependence in the relation between $b$-quark  $\overline{\rm MS}$ and pole mass and in the relation between the pole and $1S$ mass.
\end{abstract}


\section{Introduction}
Comparison of theoretical and experimental moments $M_N$, defined by
\begin{gather}
M_N= \int\frac{\mathrm{d}s}{s^{N+1}} \,R(s) \quad ,\quad N=1,2,3\ldots
\\
R(s)=\frac{\sigma\left(e^+e^-\to{\rm hadrons}\right)}{\sigma_{pt}}\quad,\quad
\sigma_{pt}=\frac{4\pi\alpha^2}{3s}~,
\end{gather}
provides a method for determining the $\overline{\rm MS}$ quark masses \cite{novikov}.
This method, combined with  recent BES data \cite{bes} (particularly in the charm threshold region) and
$\mathcal{O}\left(\alpha_s^2\right)$ (mass-dependent) perturbative expressions for the moments $M_N$ and for $R(s)$ in the continuum
region \cite{CKS}, has resulted in precision determinations of the $\overline{\rm MS}$ charm and bottom quark masses \cite{KS}.

The $\overline{\rm MS}$ $b$-quark mass determined in \cite{KS} contains three major sources of uncertainty;
\begin{enumerate}
\item experimental values of resonance and threshold contributions within $R(s)$,
\item uncertainty in $\alpha_s\left(M_Z\right)$, which enters both the perturbative series for $M_N$ and the QCD continuum contribution
to experimental moments,
\item and theoretical uncertainty associated with renormalization scale dependence within the $M_N$ perturbative
series.
\end{enumerate}
In the $b$-mass estimates of Ref.\ \cite{KS}, these three sources of uncertainty play different roles as $N$ varies, but are
generally comparable in magnitude. Thus as experimental information becomes more precise, the theoretical uncertainty devolving from
renormalization-scale dependence will become increasingly significant, and without theoretical improvement, will be the limiting factor
in the precision of $b$-mass estimates.

Techniques for substantially decreasing  renormalization scale dependence have been developed and applied to a number of
$\overline{{\rm MS}}$ perturbative
processes including semileptonic $b$ decays, light-quark contributions to $R(s)$, and Higgs decays \cite{us}.
These techniques use the appropriate renormalization-group (RG) equation for each process to determine and resum all logarithmic
contributions in the perturbation series that are explicitly determined by the RG equation.  In this paper we extend and apply such
techniques to the perturbative series for the $b$-quark contributions to $M_N$,  effectively eliminating their
renormalization-scale dependence as a source of theoretical uncertainty. It should be noted that Ref.\ \cite{KS} does not attempt to
estimate the uncertainty associated with such higher-order  contributions, so the uncertainties quoted in Ref.\ \cite{KS} may be
underestimated.

In Section \ref{scale_sec}, we develop an analysis of the $\mathcal{O}(25\,{\rm MeV})$ residual renormalization-scale dependence
characterizing the extraction of the $\overline{\rm MS}$ $b$-quark mass from the first four moments of $R(s)$.  In Section
\ref{resum_sec}, we demonstrate how this scale dependence is essentially eliminated  upon   incorporating the closed-form summation of
leading and successively-subleading logarithms within the perturbative series for $N=\{1,2,3,4\}$ moments of $b$-quark
contributions to $R(s)$.  This procedure is also shown to reduce theoretical uncertainties associated with the choice of $N$, as well
as leading to a modest ($14\,{\rm MeV}$) elevation in the central value for $m_b\left(m_b\right)$.
Finally, by assuming power-law growth in the (RG-undetermined) perturbative contributions, the effect of the (unknown) next-order
perturbative contributions is estimated.  These estimated next-order contributions are decreased by the application of resummation
techniques, providing a further reduction in a source of theoretical uncertainty.

Renormalization-scale dependence is also shown to exist as a source of uncertainty in the known perturbative expressions relating
the $b$-quark $\overline{\rm MS}$ mass to its corresponding pole mass \cite{pole} and  $1S$ mass \cite{1S,hoang}.
In Section \ref{pole_sec} we explore the scale dependence inherent in the perturbative series relating $\overline{\rm MS}$ and pole
$b$-quark masses.  In Section \ref{resum_pole_sec}, we demonstrate how this uncertainty is resolved by a renormalization-group
resummation of this series similar to that of Section \ref{resum_sec}.  Finally, in Section \ref{1S_sec} we discuss the reduction of
scale uncertainty via comparable renormalization-group resummation of the relationship between the $1S$ and pole $b$-quark masses.

\section{Residual Scale Uncertainty of the $\overline{\rm MS}$ $b$-quark Mass}
\label{scale_sec}
Following Ref.\ \cite{novikov}, K\"uhn and Steinhauser \cite{KS} express the running $b$-quark $\overline{{\rm MS}}$ mass
$m_b(\mu)$ from moments of the $b$-quark contribution to
the experimentally determined electron-positron-annihilation ratio
\begin{equation}
M_N^{exp}\equiv \int\frac{\mathrm{d}s}{s^{N+1}} \,R_b(s)
\label{v2.1}
\end{equation}
and its ($\overline{\rm MS}$) field-theoretical analogue
\begin{equation}
M_N^{th}=\left(\frac{1}{4m_b^2(\mu)}\right)^N S_N(\mu)~,
\label{v2.2}
\end{equation}
with the perturbative series $S_N$ related to $\Pi_b\left(q^2\right)$, the $b$-quark contribution to the vector-current correlation
function, via
\begin{equation}
\Pi_b\left(q^2\right)=\frac{1}{12\pi^2}\sum_{N} \left(\frac{q^2}{4m_b^2(\mu)}\right)^NS_N(\mu)~.
\label{v2.3}
\end{equation}
If one equates the experimental and theoretical moments, one finds that
\begin{equation}
m_b(\mu)=\frac{1}{2}\left(\frac{S_N(\mu)}{M_N^{exp}}\right)^{\frac{1}{2N}}~.
\label{v2.4}
\end{equation}
The series $S_N$ is a perturbative series in the QCD couplant $x(\mu)=\alpha_s(\mu)/\pi$, where $\mu$
is the renormalization scale characterizing $M_N^{th}$:
\begin{equation}
S_N(\mu)=\sum_{j=0}^\infty\sum_{k=0}^jT_{j,k}^{(N)} x^j(\mu)
\log^k{\left(\frac{\mu^2}{m_b^2(\mu)}\right)}~,
\label{v2.5}
\end{equation}
where the $j=\{0,1,2\}$ [{\it i.e.,} up to three-loop] $\overline{\rm MS}$ coefficients of this series
\cite{CKS} (summarized in Table 6 of
Ref.\cite{KS}),  are tabulated in Table \ref{vtab1}.\footnote{Our coefficients
$T_{j,k}^{(N)}$ are related to those of
Ref. \cite{KS}'s Table 6 by $T_{j,k}^{(N)}=(-1)^kc_n^{(j,k)}/4$.  Division by 4 is a consequence of
$Q_b^2=Q_c^2/4$.  The alternation in sign follows from the argument of our logarithm in
Eq.\ (\protect\ref{v2.5})
being the inverse of that for the logarithm in the series $\bar C_N$ of Ref.\ \cite{KS}.}
Since $S_N$ has  logarithmic dependence on $m_b(\mu)$, Eq.\ (\ref{v2.4}) represents an implicit equation that must be solved
numerically to determine $m_b(\mu)$.

\begin{table}
\centering
\begin{tabular}{||c|c|c|c|c||}     \hline\hline
 & $N=1$ & $N=2$ & $N=3$ & $N=4$ \\ \hline\hline
$T_{0,0}^{(N)}$  & 0.2667 & 0.1143 & 0.06772 & 0.04618 \\ \hline
$T_{1,0}^{(N)}$  & 0.6387 & 0.2774 & 0.1298 & 0.05775 \\ \hline
$T_{1,1}^{(N)}$  & -0.5333 & -0.4571 & -0.4062 & -0.3694 \\ \hline
$T_{2,0}^{(N)}$  & 0.7898 & 0.8080 & 0.5169 & 0.3051 \\ \hline
$T_{2,1}^{(N)}$  & -0.8606 & -1.2610 & -1.11454 & -0.8682 \\ \hline
$T_{2,2}^{(N)}$  & 0.0222 & 0.4762 & 0.8296 & 1.1236 \\ \hline\hline
\end{tabular}
\caption{Five-flavour ($n_f=5$) $\overline{\rm MS}$ series coefficients for $S_N$.}
\label{vtab1}
\end{table}

The theoretical
expression (\ref{v2.2}) for the moments $M_N$ is formally independent of the renormalization scale $\mu$, as expected for
this physically-observable quantity.\footnote{The experimental values for $M_N$ are tabulated in Table 7 of Ref.\ \cite{KS}}
Thus
the requirement $0=dM_N^{th} / d\mu^2$ leads to the following renormalization-group equation for the series $S_N (\mu)$:
\begin{equation}
0 = \left[ \left( 1 - 2\gamma(x) \right)\frac{\partial}{\partial L} + \beta(x) \frac{\partial}{\partial x} - 2 N \gamma(x) \right] S_N
[x,L] \label{v2.6}
\end{equation}
where $n_f = 5$ and\footnote{See Refs.\ \protect\cite{beta1,beta2} and Refs.\ \protect\cite{gamma1,gamma2} for the coefficients in
$\beta(x)$ and $\gamma(x)$, respectively.}
{\allowdisplaybreaks
\begin{gather}
L(\mu) \equiv \log \left( \mu^2 / m_b^2 (\mu) \right),
\label{v2.7}
\\
S_N (\mu)  =   S_N \left[ x(\mu), L(\mu) \right] =  \sum_{j=0}^\infty \sum_{k=0}^j T_{j,k}^{(N)} x^j L^k,
\label{v2.8}
\\
\gamma\left( x\right)  =  \frac{\mu^2}{m_b (\mu)} \frac{d m_b (\mu)}{d\mu^2} = - x \left[ 1 + \sum_{n=1}^\infty \gamma_n x^n \right],
\label{v2.9}
 \\
 \gamma_1 = \frac{253}{72}, ~ \gamma_2 = 7.41986 ,
\\
\beta \left( x\right)  =  \mu^2 \frac{dx(\mu)}{d\mu^2} = - x^2 \sum_{n=0}^\infty \beta_n x^n,
\label{v2.10}\\
 \beta_0 = \frac{23}{12},  \; \beta_1 = \frac{29}{12}, \; \beta_2 = 9769/3456  ~.
\end{gather}
}
For example, one finds from Eq.\ (\ref{v2.6}) that the coefficients $T_{1,1}^{(N)}$, $T_{2,2}^{(N)}$ and $T_{2,1}^{(N)}$ satisfy the relations
\begin{gather}
T_{1,1}^{(N)} = -2 N T_{0,0}^{(N)},
\label{v2.11}
\\
T_{2,2}^{(N)} = N\left(2N - \beta_0\right) T_{0,0}^{(N)},
\label{v2.12}
\\
T_{2,1}^{(N)} = \left(\beta_0 - 2N\right) T_{1,0}^{(N)} - 2N \left(\gamma_1 - 2\right) T_{0,0}^{(N)},
\label{v2.13}
\end{gather}
consistent (modulo round-off errors) with the entries in Table \ref{vtab1}.

We are interested in exploring both the residual renormalization scale dependence and the $N$-dependence of the $\overline{MS}$
benchmark mass $m_b \left(m_b\right)$ extracted in Ref. \cite{KS}, since each such dependence is a source of theoretical uncertainty.
As in Ref. \cite{KS}, the series (\ref{v2.5}) for $S_N (\mu)$ is truncated after its (known) $j=2$ terms. Such truncation necessarily
becomes a source of residual $\mu$ dependence.  Since we are focusing {\em only} on theoretical uncertainties arising from such scale
dependence, we assume that $x(\mu)$ four-loop evolves from its Ref. \cite{KS} benchmark value $x\left(M_z\right) = 0.11800/\pi$ to
$x(10 \; {\rm GeV}) = 0.056732$, and disregard theoretical uncertainty associated with $x\left(M_z\right)$
[and hence $x(10\,{\rm GeV})$]. The preferred
renormalization scale in Ref. \cite{KS} for extracting $m_b \left(m_b\right)$ is $10 \; {\rm GeV}$, and as in Ref.\ \cite{KS}, the scale
dependence we consider is over the range $5 \; {\rm GeV} \leq \mu \leq 15 \; {\rm GeV}$.  The extraction
of $m_b\left(m_b\right)$ occurs first by numerical solution of  Eq.\ (\ref{v2.2}) to obtain $m_b (\mu)$, and then by
evolving $m_b(\mu)$ downward via Eq.\ (\ref{v2.9}) to the point where $\mu = m_b (\mu)\equiv m_b\left(m_b\right)$.

In Table \ref{vtab2}, we list the above-described extractions of $m_b (\mu)$ for $\mu = 5$, $10$ and $15 \; {\rm GeV}$, as obtained
from each of the $N = \{1, 2, 3, 4 \}$ moments using values for $M_N^{\exp}$ from Table 7 of Ref. \cite{KS}.
\footnote{Table \protect\ref{vtab2} numbers for $\mu
= 10 \; {\rm GeV}$ are in agreement with those of Ref. \cite{KS}'s Table 8.}

\begin{table}
\centering
\begin{tabular}{||c|c|c|c|c|c|c||} \hline\hline
$\mu$ & $N=1$ & $N=2$ & $N=3$ & $N=4$ & $\overline{m_b} (\mu)$ & $\sigma_N$\\ \hline\hline
5  & 4.0848 & 4.0799 & 4.0751 & 4.0707 & 4.0776 & 0.0053 \\ \hline
10  & 3.6652 & \bf{3.6509} & 3.6407 & 3.6551 & 3.6530 & 0.0088 \\ \hline
15  & 3.4708 & 3.4457 & 3.4445 & 3.5290 & 3.4725 & 0.034 \\ \hline\hline
\end{tabular}
\caption{$m_b^{(N)} (\mu)$ as extracted from (\ref{v2.4}) via $S_N (\mu)$ truncated after three-loop terms.  All entries are in ${\rm
GeV}$. The bold-face entry $[N = 2, \; \mu = 10\,{\rm GeV}]$ corresponds to that preferred in Ref.~\cite{KS} to generate
$m_b\left(m_b\right)$.} \label{vtab2}
\end{table}

\smallskip

The column $\overline{m_b}(\mu)$ of Table \ref{vtab2} is just the average of $m_b (\mu)$ taken over the first four moments.  The rms
spread of values over the first four moments is
\begin{equation}
\sigma_N = \frac{1}{2} \left[ \sum_{N=1}^4 \left(m_b^{(N)} (\mu) - \overline{m^{(N)}_b} (\mu) \right)^2 \right]^{1/2}.
\label{v2.14}
\end{equation}
One sees immediately from the table that this rms spread of values for $m_b(\mu)$ increases dramatically with $\mu$, indicative of residual scale dependence.
In other words, the error associated with different choices of $N$ is itself a scale dependent quantity.

\begin{figure}[hbt]
\centering
\includegraphics[scale=0.5]{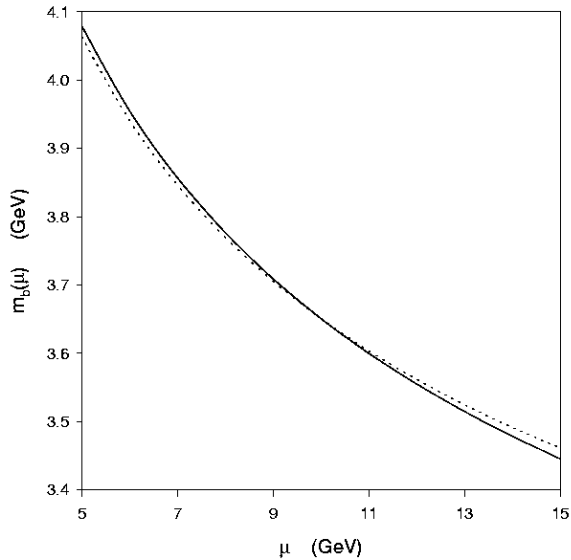}
\caption{
Renormalization-scale ($\mu$) dependence of  $m_b(\mu)$  extracted via Eq.\ (\protect\ref{v2.4}) (solid curve) for the $N=2$ moment
compared with the RG evolution of $m_b(\mu)$ (broken curve). For RG evolution,  $m_b(10\,{\rm GeV})$ is used as a reference scale, and
hence the two curves intersect at $\mu=10\,{\rm GeV}$. }
\label{fig1}
\end{figure}

\begin{figure}[hbt]
\centering
\includegraphics[scale=0.5]{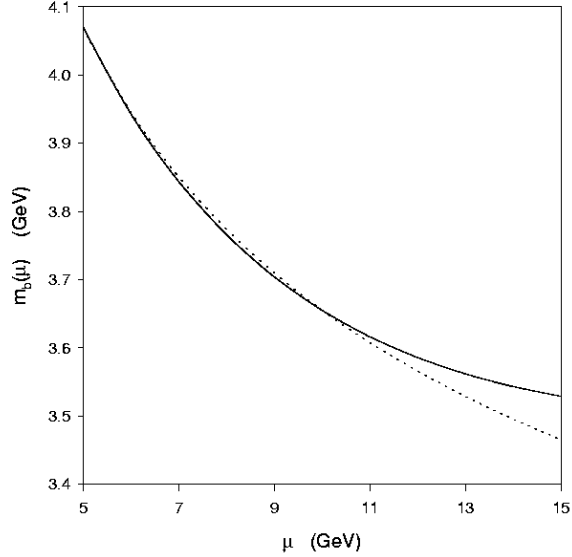}
\caption{
Renormalization-scale ($\mu$) dependence of  $m_b(\mu)$  extracted via Eq.\ (\protect\ref{v2.4}) (solid curve) for the $N=4$ moment
compared with the RG evolution of $m_b(\mu)$ (broken curve). For RG evolution,  $m_b(10\,{\rm GeV})$ is used as a reference scale, and
hence the two curves intersect at $\mu=10\,{\rm GeV}$. }
\label{fig2}
\end{figure}

If Table \ref{vtab2} represents a valid determination of the quark mass, then for a fixed $N$, the variation of $m_b(\mu)$
with $\mu$ should
conform with the RG evolution equation (\ref{v2.9}).
However, the $\mu$-dependence of $m_b (\mu)$ as extracted via Eq.\ (\ref{v2.2}) is not fully consistent with such evolution.
In Figure \ref{fig1}, we have plotted the $\mu$-dependence of such extracted values for the $N = 2$ case against the (three loop)
RG-evolution following from Eq.\ (\ref{v2.9}) for values of $\mu$ between $5\,{\rm GeV}$  and  $15\,{\rm GeV}$  To facilitate this
comparison, we evolve from the same value $m_b (10\,{\rm GeV})$ as extracted in Table \ref{vtab2} (in bold).  The figure clearly shows
a deviation by the $\mu$-dependence extracted from Eq.\ (\ref{v2.2}) from that anticipated from RG-evolution.  Moreover, this deviation
becomes progressively pronounced with increased $N$.  For $N = 4$, Figure \ref{fig2} shows the plot of extracted versus RG-evolved
$\mu$ dependence, which exhibits a substantially larger deviation for large $\mu$, though a somewhat better fit between $\mu = 5\,{\rm
GeV}$ and $\mu = 10 \; {\rm GeV}$.
The deviation exhibited in the Figures of the extracted $b$-quark mass from the behaviour expected from RG evolution is significant
since it is much larger than the effects associated with a change to one-loop higher (or lower) in the RG-evolution curve.

These identifiable remaining scale dependences, both horizontal ($N$-dependence) and vertical (deviation from RG-evolution),
necessarily percolate into estimates of $m_b \left(m_b\right)$.  In Table \ref{vtab3} we display values of $m_b \left(m_b\right)$
obtained by evolving Eq.\ (\ref{v2.2}) values $m_b (\mu)$ extracted from Eq.\ (\ref{v2.4}) at the indicated scale $\mu$.  The values
for $\mu = 10 \; {\rm GeV}$, including the preferred $N = 2$ value, differ inconsequentially (they are $2 \; {\rm MeV}$ larger) from
the central values displayed in the erratum to Ref. \cite{KS}, providing a check on our calculation.  The column $\overline{m_b}
\left(m_b\right)$ of Table \ref{vtab3} is just the average over $N$ of $m_b \left(m_b\right)$ values, as evolved from the indicated
choice of $\mu$.

\begin{table}
\centering
\begin{tabular}{||c|c|c|c|c|c|c||} \hline\hline $\mu$ & $N=1$ & $N=2$ & $N=3$ & $N=4$ &
$\overline{m_b} \left(m_b\right)$ & $\sigma_N$\\ \hline\hline 5  & 4.2116 & 4.2073 & 4.2031 & 4.1992 & 4.2053 & 0.0046 \\ \hline 10  &
4.2067 & \bf{4.1928} & 4.1830 & 4.1969 & 4.1948 & 0.0085 \\ \hline 15  & 4.2025 & 4.1768 & 4.1755 & 4.2620 & 4.2042 & 0.035 \\
\hline\hline $\sigma_\mu$  & 0.005 & 0.015 & 0.014 & 0.033 & 0.0052 &  \\ \hline\hline
\end{tabular}
\caption{$m_b^{(N)} \left(m_b\right)$ as RG-evolved from $m_b^{(N)}$ values listed in Table \ref{vtab2} ({\it i.e.}, via 3-loop-truncated series $S_N$).
All entries are in ${\rm GeV}$.  The bold-faced entry $[N=2, \mu=10]$ corresponds to preferred choices in Ref.\ \cite{KS}.}
\label{vtab3}
\end{table}

\smallskip

Similarly, the rms spread over $N$ is just
\begin{equation}
\sigma_N = \frac{1}{2} \left[ \sum_{N=1}^4 \left\{m_b^{(N)} \left(m_b\right) - \overline{m_b} \left(m_b\right) \right\}^2
\right]^{1/2}, \label{v2.15}
\end{equation}
and the scale uncertainty ($\sigma_\mu$) is just half the difference between the maximum and minimum value of $m_b \left(m_b\right)$
 over a given column of the table.  Thus $\sigma_N$ is a measure of horizontal ($N$-dependence) uncertainty, and $\sigma_\mu$ is
 a measure of
vertical ($\mu-)$ renormalization-scale uncertainty.
For the $N=2$ $\mu=10\,{\rm GeV}$ preferred case \cite{KS}, both of these uncertainties
are indicative of  overall
$24 \; {\rm MeV}$ theoretical uncertainties in $m_b\left(m_b\right)$ that devolve ultimately from residual scale
dependence in truncating the series $S_N$.

Another source of theoretical uncertainty is the effect of next- and higher-order contributions in the perturbative series used to
determine $m_b(\mu)$. Such effects  were not considered in the error analysis of \cite{KS}.  The RG equation is capable of determining
$T^{(N)}_{3,n}$ for $n=\{1,2,3\}$
{\allowdisplaybreaks
\begin{gather}
-T^{(N)}_{3,3}=\left(\frac{2}{3}N\beta_0^2-2\beta_0 N^2+\frac{4}{3}N^3\right) T^{(N)}_{0,0}\\
T^{(N)}_{3,2}=\left(-8N^2+6N\beta_0-N\beta_1+4N^2\gamma_1-2N\beta_0\gamma_1 \right)T^{(N)}_{0,0}+
\left(-3N\beta_0+\beta_0^2+2N^2 \right)T^{(N)}_{1,0}\\
-T^{(N)}_{3,1}=\left(8N+2N\gamma_2-8N\gamma_1\right)T^{(N)}_{0,0}
+\left(2N\gamma_1+2\beta_0-\beta_1-4N\right)T^{(N)}_{1,0}
+\left(-2\beta_0+2N\right)T^{(N)}_{2,0} \quad ,
\end{gather}
}
leaving only $T_{3,0}^{(N)}$ undetermined.  However, we can {\em estimate} the effect of these next-order contributions by assuming
that the approximate power-law growth exhibited by $T_{0,0}^{(N)}$, $T_{1,0}^{(N)}$, and $T_{2,0}^{(N)}$ continues at next-order,
resulting in the estimates
\begin{equation}
T_{3,0}^{(1)}=3.64,~T_{3,0}^{(2)}=2.15,~T_{3,0}^{(3)}=1.42,~T_{3,0}^{(4)}=0.784\quad .
\label{T30_eq}
\end{equation}
In addition to inclusion of the $T_{3,n}$, it is necessary to include the next-order ($n_f=5$) terms
\begin{equation}
\beta_3=18.8522,~\gamma_3=11.0343
\end{equation}
in the evolution of the running coupling and mass.
These higher-order terms lead to an additional theoretical uncertainty of approximately $20\,{\rm MeV}$ for the $N=2$ benchmark
$m_b\left(m_b\right)$ value, comparable to the  renormalization-scale uncertainty in Table \ref{vtab3}.  By comparison, the
uncertainty in $m_b\left(m_b\right)$ arising from the experimental inputs ($M_2^{exp}$ \cite{KS} and $\alpha_s\left(M_Z\right)$
\cite{pdg}) is approximately $40\,{\rm MeV}$, and hence the theoretical and experimental uncertainties are of comparable magnitude.

\section{Optimal RG-Improvement of the $\overline{{\rm MS}}$ $b$-quark Mass}
\label{resum_sec}
As noted in the previous section, the higher order $S_N (\mu)$ series coefficients $T_{1,1}^{(N)}$ and $T_{2,2}^{(N)}$
are determined via the RG-equation (\ref{v2.6}) from the leading series coefficient $T_{0,0}^{(N)}$.  Similarly, the
three-loop coefficient $T_{2,1}^{(N)}$ can be obtained via Eq.\ (\ref{v2.13}) from the one and two-loop series terms
$T_{0,0}^{(N)}$ and $T_{1,0}^{(N)}$.  In fact, the RG-equation (\ref{v2.6}) is much more powerful than any  use we have made
of it so far.  Given the calculated values of $T_{0,0}^{(N)}$, $T_{1,0}^{(N)}$ and $T_{2,0}^{(N)}$, one can determine
respectively {\it every} leading-logarithm (LL) coefficient $T_{j,j}^{(N)}$, {\it every} next-to-leading-logarithm (NLL)
coefficient $T_{j, j-1}^{(N)}$, and {\it every} next-to-next-to-leading logarithm (NNLL) coefficient $T_{j, j-2}^{(N)}$ in the
series expansion (\ref{v2.5}) for $S_N (\mu)$.

The procedure of optimal RG improvement \cite{us} involves the summation to all orders of leading and progressively subleading
logarithms within a series, a process that has been seen to reduce significantly the renormalization scale dependence in a wide
variety of processes \cite{us}.  For the case at hand, we wish to include every RG-accessible coefficient $T_{j, k}^{(N)}$ in the
series $S_N(\mu)$ in order to extract via Eq.\ (\ref{v2.2}) an $\overline{{\rm MS}}$ $b$-quark mass $m_b \left(m_b\right)$ that is free
(or nearly so) of the residual scale dependence evident in Table \ref{vtab3}.  To do this, we first organize the series (\ref{v2.5})
as follows:
\begin{equation}
S_N [x,L] = \sum_{n=0}^\infty x^n S_n^{(N)} (xL),
\label{v3.16}
\end{equation}
where
\begin{equation}
S_n^{(N)} (u) = \sum_{k=n}^\infty T_{k, k-n}^{(N)} u^{k-n}
\label{v3.17}
\end{equation}
with $u = xL$ amounts to an LL summation when $n = 0$, an NLL summation when $n = 1$, and an NNLL summation when $n = 2$.  If
we substitute
Eq.\ (\ref{v3.16}) into Eq.\ (\ref{v2.6}),
we generate a succession of first-order differential equations for these summations:
{\allowdisplaybreaks
\begin{gather}
\left[ \left(1 - \beta_0 u\right) \frac{d}{du} + 2N \right] S_0^{(N)} (u) = 0,
\label{v3.18}
\\
\left[ \left(1 - \beta_0 u\right) \frac{d}{du}  +  \left(2N - \beta_0\right) \right] S_1^{(N)} (u)
=  \left[ \left(\beta_1 u - 2\right) \frac{d}{du} - 2N \gamma_1 \right] S_0^{(N)} (u),
\label{v3.19}
\\
\begin{split}
0=&-\left[ \left(1 - \beta_0 u\right) \frac{d}{du} + \left(2N - 2\beta_0\right) \right] S_2^{(N)} (u)
\\*
&
+  \left[ \left(\beta_1 u - 2\right) \frac{d}{du} + \left(\beta_1 - 2N \gamma_1 \right) \right] S_1^{(N)} (u)
+ \left[ \left( \beta_2 u - 2 \gamma_1\right) \frac{d}{du} - 2 N \gamma_2 \right] S_0^{(N)} (u).
\end{split}
\label{v3.20}
\end{gather}
}
Eq.\ (\ref{v3.17}) provides the initial conditions $S_0^{(N)}(0) = T_{0,0}^{(N)}$,
$S_1^{(N)}(0) = T_{1,0}^{(N)}$, $S_2^{(N)}(0) = T_{2,0}^{(N)}$, the set of all known $T_{k,0}^{(N)}$ coefficients of the series $S_N
(\mu)$ [see Table \ref{vtab1}].  With these initial conditions, Eqs.\ (\ref{v3.18})--(\ref{v3.20}) can be successively solved, and the
optimally RG-improved series is found to be
\begin{equation}
S^\Sigma_N [xL] = S_0^{(N)} (xL) + x S_1^{(N)} (xL) + x^2 S_2^{(N)} (xL).
\label{v3.21}
\end{equation}
For
\begin{equation}
w \equiv 1 - \beta_0 xL, \; \; A \equiv -2N / \beta_0,
\label{v3.22}
\end{equation}
we find from Eqs.\ (\ref{v3.18}) and (\ref{v3.19}) that
\begin{equation}
S_0^{(N)} (xL) = T_{0,0}^{(N)} w^{-A},
\label{v3.23}
\end{equation}
and
\begin{equation}
S_1^{(N)}(xL) = B w^{-A} + \left[ T_{1,0}^{(N)} - B + C \log w\right] w^{-A-1},
\label{v3.24}
\end{equation}
with
\begin{equation}
B \equiv \left(\beta_1 A + 2N \gamma_1\right) T_{0,0}^{(N)} / \beta_0,~
C \equiv \left(2 \beta_0 - \beta_1\right) A T_{0,0}^{(N)} / \beta_0.
\label{v3.25}
\end{equation}
Substituting Eqs.\ (\ref{v3.23}) and (\ref{v3.24}) into Eq.\ (\ref{v3.20}) we obtain the solution
\begin{equation}
\begin{split}
S_2^{(N)}(xL) =& \frac{D}{2} w^{-A} + w^{-A-1} \left[ E - F + F \log (w) \right]
\\
&+ w^{-A-2} \left[ T_{2,0}^{(N)} - \frac{D}{2} - E + F + G \log (w) + \frac{H}{2} \log^2 w\right],
\end{split}
\label{v3.26 }
\end{equation}
with
{\allowdisplaybreaks
\begin{gather}
D = \left[ \beta_1 AB + \beta_2 AT_{0,0}^{(N)} - \left(\beta_1 - 2N \gamma_1\right) B+2N \gamma_2 T_{0,0}^{(N)} \right] /
\beta_0, \label{v.3.27}
\\
\begin{split}
E  = &\Biggl[ \left(2\beta_0 - \beta_1\right) AB + \beta_1 \left[ (1+A) \left(T_{1,0}^{(N)} - B\right) - C \right]\Biggr.
 \\*
& + \Biggl.\left(2 \gamma_1 \beta_0 - \beta_2\right)AT_{0,0}^{(N)} +   \left(B - T_{1,0}^{(N)}\right) \left(\beta_1 - 2N
\gamma_1\right)\Biggr]/\beta_0,
\end{split}
\label{v3.28}
\\
F = C\left(A\beta_1 + 2N \gamma_1\right) / \beta_0,
\label{v3.29}
\\
G = \left[ (1 + A)\left(T_{1,0}^{(N)} - B\right) - C \right] \left(2\beta_0 - \beta_1\right) / \beta_0,
\label{v3.30}
\\
H = \left(2 \beta_0 - \beta_1\right)(1+A) C / \beta_0.
\label{v3.31}
\end{gather}

The extraction of $m_b \left(m_b\right)$ now proceeds analogously to that in the previous section, except that the series $S_N (\mu)$ is
now in the optimally RG-improved form (\ref{v3.21}), rather than the truncation
\begin{equation}
S_N = \sum_{j=0}^2 \sum_{k=0}^j T_{j,k}^{(N)} x^j L^k
\label{v3.32}
\end{equation}
of Eq.\ (\ref{v2.5}) to one-, two-, and three-loop contributions utilized in Section 2 (and employed in Ref. \cite{KS}).
For a given choice of renormalization
scale $\mu$, the values of $m_b(\mu)$ we extract from Eqs.\ (\ref{v2.2}) and (\ref{v3.21}) are tabulated in Table \ref{vtab4}.
As before, the average $\overline{m_b} (\mu)$ is over $N = \{1,2,3,4\}$ values of $m_b (\mu)$, and the rms spread over
these values is given by Eq.\ (\ref{v2.14}).
These spreads are seen to be significantly less than those of Table \ref{vtab2}.  We thus see that the extracted
values for $m_b (\mu)$ from different values of $N$ are in much better agreement when $S_N$ is RG-improved.
Moreover, the substantial increase of $\sigma_N$ with $\mu$ characterizing Table \ref{vtab2} (and indicative of
residual scale dependence)
does {\em not} occur in
Table \ref{vtab4}.  In Table \ref{vtab4}, $\sigma_N$ is essentially static at $4$--$5 \; {\rm MeV}$ for $\mu$ between $5$ and $15 \;
{\rm GeV}$, corresponding to a small fixed theoretical uncertainty associated with the choice for $N$.  Thus, RG-improvement is
seen to disentangle (vertical) scale-uncertainties from (horizontal) $N$-uncertainties, as well as to reduce the magnitude of
such $N$-uncertainties.

\begin{table}
\centering
\begin{tabular}{||c|c|c|c|c|c|c||} \hline\hline
$\mu$ & $N=1$ & $N=2$ & $N=3$ & $N=4$ & $\overline{m_b} (\mu)$ & $\sigma_N$\\ \hline\hline
5  & 4.0851 & 4.0797 & 4.0750 & 4.0717 & 4.0779 & 0.0050 \\ \hline
10  & 3.6698 & 3.6653 & 3.6610 & 3.6582 & 3.6636 & 0.0044 \\ \hline
15  & 3.4793 & 3.4751 & 3.4712 & 3.4687 & 3.4736 & 0.0040 \\ \hline\hline
\end{tabular}
\caption{$m_b^{(N)} (\mu)$ as extracted from optimally RG-improved $S_N (\mu)$.   All entries are in ${\rm GeV}$.  The average $m_b(\mu)$
and the rms spread $\sigma_N$ over the four values of $N$ are calculated as in Table \ref{vtab2}.  All entries are in ${\rm GeV}$.}
\label{vtab4}
\end{table}

In contrast to Figs.\ \ref{fig1} and \ref{fig2}, the Table \ref{vtab4} values of $m_b (\mu)$ extracted from Eq.\ (\ref{v2.4}) via the
resummed $S_N^\Sigma$ of Eq.\ (\ref{v3.21}) are fully consistent  with the RG-evolution equation (\ref{v2.9}).
For $N = 2$ Figure \ref{fig3}
plots both extracted values of $m_b(\mu)$  for $\mu$ between $5\, {\rm GeV}$ and $15 \; {\rm GeV}$, as well as the values evolved [via
Eq.\ (\ref{v2.9})] directly from the extracted value $m_b^{(2)} (10 \; {\rm GeV}) = 3.6653 \; {\rm GeV}$.
The points coincide to
within the visual resolution of the figure, indicative of purely RG scale-dependence for values of $m_b(\mu)$ extracted
via Eq.\ (\ref{v3.21}).\footnote{The difference between extracted and evolved values at $\mu = 15 \; {\rm GeV}$ is
less than $10^{-4} \; {\rm GeV}$, based on identical starting values at $\mu = 10 \; {\rm GeV}$. This difference is
characteristic of the effect generated by including next-order terms in the RG-evolution equation.}

\begin{figure}[hbt]
\centering
\includegraphics[scale=0.5]{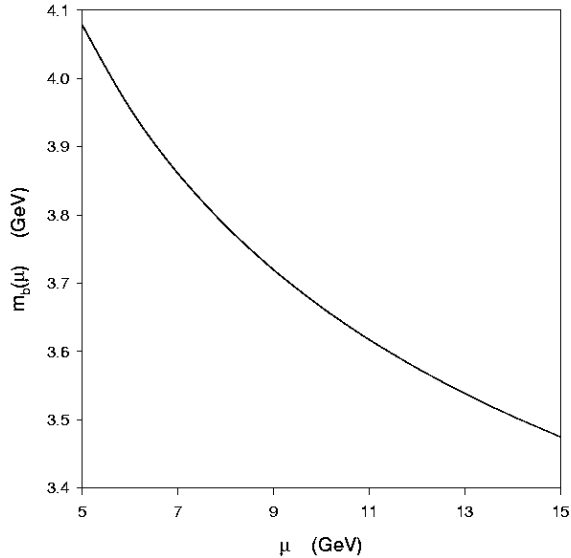}
\caption{
Renormalization-scale ($\mu$) dependence of  $m_b(\mu)$  extracted via substitution of  the resummed quantity $S_N^\Sigma$ into
(\protect\ref{v2.4}) (solid curve) for the $N=2$ moment compared with the RG evolution of $m_b(\mu)$ (broken curve). For RG evolution,
$m_b(10\,{\rm GeV})$ is used as a reference scale, and hence the two curves intersect at $\mu=10\,{\rm GeV}$.
The two curves overlap completely within the resolution of the figure.  Similar overlap occurs for all other  moments considered
({\it i.e.} $N=1,3,4$).  }
\label{fig3}
\end{figure}

The absence of any additional residual scale dependence
naturally carries over to the RG-evolution of extracted $m_b(\mu)$ to $m_b\left(m_b\right)$. In Table \ref{vtab5} we list values of
$m_b \left(m_b\right)$ obtained by evolution of the value  $m_b (\mu)$ extracted at the scale $\mu$ for the indicated moments $N$.
RG-improvement is seen in Table \ref{vtab4} to virtually eliminate the scale uncertainty ($\sigma_\mu$) evident in Table \ref{vtab3}.
The only theoretical uncertainty still evident is horizontal, the $\sigma_N$ associated with different choices of $N$, and this
uncertainty is both small ($\mathcal O(4 \; {\rm MeV})$) and static as $\mu$ varies from $5$ to $15 \; {\rm GeV}$.  For
Ref.\ \cite{KS}'s phenomenologically motivated choice $\mu = 10 \; {\rm GeV}$, $\sigma_N$ for the RG improved case (Table \ref{vtab5})
is less than half the value for $\sigma_N$ when $S_N$ is truncated (Table \ref{vtab3}).

\begin{table}
\centering
\begin{tabular}{||c|c|c|c|c|c|c||} \hline\hline
$\mu$ & $N=1$ & $N=2$ & $N=3$ & $N=4$ & $\overline{m_b} \left(m_b\right)$ & $\sigma_N$\\ \hline\hline
5  & 4.2118 & 4.2071 & 4.2030 & 4.2002 & 4.2055 & 0.0044 \\ \hline
10  & 4.2112 & 4.2068 & 4.2027 & 4.2000 & 4.2052 & 0.0042 \\ \hline
15  & 4.2111 & 4.2069 & 4.2028 & 4.2003 & 4.2053 & 0.0041 \\ \hline\hline
$\sigma_\mu$  & 0.0007 & 0.0003 & 0.0003 & 0.0003 & 0.0003 &  \\ \hline\hline
\end{tabular}
\caption{$m_b^{(N)} \left(m_b\right)$ as RG-evolved from $m_b^{(N)} (\mu)$ values listed in Table \ref{vtab4} ({\it i.e.}, via the
RG-improved series $S^\Sigma_N$). $\overline{m_b} \left(m_b\right)$, the average of $m_b\left(m_b\right)$ over $N$, $\sigma_N$,
the rms spread over $N$, and the scale uncertainty $\sigma_\mu$ are obtained as in Table \ref{vtab2}.  All entries are in ${\rm GeV}$.}
\label{vtab5}
\end{table}

\smallskip

This virtual elimination of residual scale dependence, as evident in Table \ref{vtab5}, also leads to a changed central value for
$m_b \left(m_b\right)$ relative to that of the erratum to Ref. \cite{KS}.  The central value quoted in the erratum
is $4.191 \; {\rm GeV}$, based on choices $\mu = 10 \; {\rm GeV}$, $N=2$.  The corresponding value in our Table \ref{vtab3} is
$4.193 \; {\rm GeV}$, and the $2 \; {\rm MeV}$ discrepancy is insignificant compared to
the erratum estimate of theoretical uncertainty $(\pm 51 \; {\rm MeV}$), or relative to Table \ref{vtab3}'s
$\pm 15 \; {\rm MeV}$ and $\pm 9 \;{\rm MeV}$ vertical and horizontal theoretical uncertainties associated with the choice of
 $\mu$ and $N$.  The corresponding $N=2$, $\mu = 10 \; {\rm GeV}$ value in Table \ref{vtab5} is 
 $4.207 \; {\rm GeV}$ with aggregate
 $4.5 \;{\rm MeV}$ horizontal and vertical uncertainties.

Thus, the incorporation of an optimally RG-improved perturbative series $S_N (\mu)$ is seen to eliminate the renormalization scale
theoretical uncertainty
and  halve the moment-dependence theoretical uncertainty, with a $14 \; {\rm MeV}$ increase in the corresponding
central value $4.207\,{\rm GeV}$.
Note the near equivalence  of this value with the $m_b \left(m_b\right) = 4.205 \; {\rm GeV}$ values found from the
$\overline{ m_b}\left(m_b\right)$ averages over the first four moments (Table \ref{vtab5}).

As in the previous section, the approximated values (\ref{T30_eq}) for $T_{3,0}^{(N)}$ can be used to estimate the effect of
higher-order perturbative corrections on the extraction of $m_b\left(m_b\right)$.  However, with the input of $T_{3,n}^{(N)}$,
it is necessary to extend (\ref{v3.21}) to include an $S_3^{(N)}(xL)$ term
\begin{equation}
S^\Sigma_N [xL] = S_0^{(N)} (xL) + x S_1^{(N)} (xL) + x^2 S_2^{(N)} (xL)+ x^3 S_3^{(N)} (xL)\quad .
\end{equation}
The expression for $S_3^{(N)}(xL)$ is determined by an extension of equations (\ref{v3.18})--(\ref{v3.20}),
and the final result
can be extracted by appropriate modifications of \cite{us}.  The resulting expression is
\begin{equation}
\begin{split}
S^{(N)}_3(xL)  =&  \frac{K}{3} w^{-A} + \left(\frac{M}{2} - \frac{N}{4}\right) w^{-A-1}
 +  \frac{N}{2} w^{-A-1} \log (w) + (P-Q + 2R) w^{-A-2}
 +  (Q-2R) w^{-A-2} \log (w)
\\
&+ R w^{-A-2} \log^2 (w)
 +  \left(-\frac{K}{3} - \frac{M}{2} + \frac{N}{4} - P + Q - 2R + T_{3,0} \right) w^{-A-3}
\\
& +  U w^{-A-3} \log (w) + \frac{V}{2} w^{-A-3} \log^2(w)
 +  \frac{Y}{3} w^{-A-3} \log^3 (w)\quad ,
\end{split}
\label{S3_decay_res}
\end{equation}
where the ($N$-dependent) coefficients are given by
{\allowdisplaybreaks
\begin{gather}
K  =  \frac{A}{\beta_0}\left(\beta_3T^{(N)}_{0,0} + B\beta_2 + D \beta_1/2\right)
 - \frac{1}{\beta_0} \left[ -2N\gamma_3T^{(N)}_{0,0} + \left(-2N\gamma_2+\beta_2\right) B + \left(-2N\gamma_1+2\beta_1\right)
\frac{D}{2}\right]
\label{K_def}
\\
\begin{split}
M  =&  \left[\left(2\gamma_2 - \frac{\beta_3}{\beta_0}\right)T^{(N)}_{0,0}
+ \left(2\gamma_1 - \frac{\beta_2}{\beta_0}\right)B + \left(2\gamma_0 - \frac{\beta_1}{\beta_0}\right)
\frac{D}{2} \right] A
\\
&+  \left[ \left(-2N\gamma_2 + \beta_2\right)\left(B-T^{(N)}_{1,0}\right) + \left(-2N\gamma_1 +
2\beta_1\right)(F-E)\right] \frac{1}{\beta_0}
\\
 &
 +  \left[ \left(T^{(N)}_{1,0} - B\right)(1+A)-C \right] \frac{\beta_2}{ \beta_0}
 +  \left[ E(1+A) - F(2+A)\right] \frac{\beta_1}{ \beta_0}
\end{split}
\\
N = \left\{ \left(A\beta_2 + 2N\gamma_2\right) C+\left[ (A-1)\beta_1 +2N\gamma_1\right] F\right\}\frac{1}{ \beta_0}
\\
\begin{split}
P  =&  \left(2\gamma_1 - \frac{\beta_2}{ \beta_0}\right) \left[ (1+A)\left(T^{(N)}_{1,0} - B\right) -
C\right]
 +  \left(2\gamma_0 - \frac{\beta_1}{ \beta_0}\right) \left[ (1+A) E - (2+A)
F\right]
\\
& - \frac{\left(-2N\gamma_1 + 2\beta_1\right)}{\beta_0}\left(T^{(N)}_{2,0} - \frac{D}{2} - E + F\right)
 -  \frac{\beta_1}{\beta_0} \left[ G-(2+A) \left(T^{(N)}_{2,0} - \frac{D}{2} - E + F\right) \right]
\end{split}
\\
Q  =  \left[ \left(2\gamma_1 - \frac{\beta_2}{ \beta_0}\right) C + \left(2\gamma_0 -
\frac{\beta_1}{\beta_0}\right) F\right] (1+A)
 -  \left[ \left(-2N\gamma_1 + 2\beta_1\right) G + (H - (2+A) G) \beta_1 \right]
\frac{1}{\beta_0}
\\
R = \left(\beta_1 A + 2N\gamma_1\right) \frac{H}{ 2\beta_0}
\\
U = \left(2\gamma_0 - \frac{\beta_1}{ \beta_0}\right) \left[ (2+A) \left(T^{(N)}_{2,0} - \frac{D}{2} - E + F\right)
-G \right]
\\
V = \left(2\gamma_0 - \frac{\beta_1}{\beta_0}\right) \left[ (2+A) G - H \right]
\\
Y = \left(2\gamma_0 - \frac{\beta_1}{ \beta_0}\right)(2+A) \frac{H}{2} \quad.
\end{gather}
}
The effect of these higher-order terms lead to an approximate $7\,{\rm MeV}$ uncertainty in the $N=2$ benchmark  $m_b\left(m_b\right)$
value.  This is a significant reduction compared with the $20\,{\rm MeV}$ uncertainty occurring for the un-summed case described
previously.  By comparison, the uncertainty in the resummed $m_b\left(m_b\right)$ arising from the experimental inputs
($M_2^{exp}$ \cite{KS} and $\alpha_s\left(M_Z\right)$ \cite{pdg}) is approximately $40\,{\rm MeV}$, and hence the resummation analysis
reduces theoretical uncertainties to a level well below the experimental uncertainties.

\section{Residual Scale Dependence of the Pole Mass}
\label{pole_sec}
The series $T(\mu) = T\left[ x(\mu), L(\mu) \right]$ relating the (RG-invariant) pole $b$-quark mass $M_b^p$ and the $\overline{{\rm
MS}}$ mass $m_b (\mu)$ is given by
\begin{equation}
M_b^p = m_b (\mu) T \left[ x(\mu), L(\mu) \right],
\label{v4.33}
\end{equation}
\begin{equation}
T[x,L] = 1 + \sum_{j=0}^\infty \sum_{k=0}^j T_{j,k} x^j L^k~,
\label{v4.34}
\end{equation}
where $x(\mu) = \alpha_s (\mu) / \pi$ and $L(\mu) = \log \left( \mu^2 / m_b^2 (\mu) \right)$
as before, and where the known series coefficients in the $L = 0$ limit are \cite{pole}
\begin{gather}
T_{1,0}  =  4/3,
\nonumber\\
T_{2,0}  =  -1.0414 \left(n_f - 1\right) + 13.4434 \;
\stackrel{n_f = 5}{\longrightarrow}
9.2778
\nonumber\\
T_{3,0}  =  0.6527 \left(n_f - 1\right)^2 - 26.655 \left(n_f - 1\right)
 +  190.595 \;
\stackrel{n_f = 5}{\longrightarrow}
94.4182
\label{v4.35}
\end{gather}
If one wishes to relate the pole mass to $m_b \left(m_b\right)$ ({\it i.e.}, to the point $\mu = m_b (\mu)$), all logarithms in
the series (\ref{v4.34}) are zero and only the coefficients $T_{j,0}$ contribute.

Since $M^p_b$ is a RG-invariant quantity, renormalization scale dependence can be studied by explicitly varying $m_b(\mu)$ through the
RG equation as $\mu$ is varied. Given the residual scale dependence implicit in any truncation of the series (\ref{v4.34}), one
can argue that the choice of scale $\mu$, generally motivated by experimental information [as
is used to determine $M_N^{\exp}$ via Eq.\
(\ref{v2.1})], should be consistently maintained. For example,  the benchmark mass $m_b \left(m_b\right)$ is obtained
in Ref.\ \cite{KS} via a
determination of $m_b (10 \; {\rm GeV})$, and then by subsequent evolution [Eq.\ (\ref{v2.9})] to the point $\mu = m_b (\mu)$.

A  measure of the residual scale-dependence implicit in the determination of the pole mass from the $\overline{{\rm MS}}$ mass
obtained via Ref. \cite{KS} methodology would be the difference between
\begin{enumerate}
\item the $\mu = 10 \;  {\rm GeV}$ pole mass obtained via Eqs.\ (\ref{v4.33}) and (\ref{v4.34}) by incorporating
within the logarithm $L(\mu)$
the value $m_b (10 \;{\rm GeV})$ actually extracted from Eq.\ (\ref{v2.4}), and
\item the $\mu = m_b \left(m_b\right)$ pole mass obtained via Eqs.\ (\ref{v4.33}) and (\ref{v4.34})  from
similar incorporation of $m_b
\left(m_b\right)$, as evolved via Eq.\ (\ref{v2.9}) from the extracted value $m_b (10 \; {\rm GeV})$.
\end{enumerate}
We emphasize that if the input values $m_b \left(m_b\right)$ and $m_b ( 10 \; {\rm GeV})$ are RG-consistent, then any discrepancy between $M_b^p$
obtained by these two procedures is a reflection of residual scale dependence arising solely from the truncation of the series (\ref{v4.34}).

To implement this comparison, we need to know the coefficients $T_{j,k}$ with $k \neq 0$ for $j = \{1, 2, 3 \}$.  Since the pole mass is an
RG-invariant,
 $d M_b^p / d\mu^2 = 0$. One then finds from Eq.\ (\ref{v4.33}) the following RG-equation for the series $T[x,L]$:
\begin{equation}
\left[ \left( 1 - 2 \gamma (x) \right) \frac{\partial}{\partial L} + \beta (x) \frac{\partial}{\partial x} + \gamma (x) \right] T [x, L] = 0
\label{v4.36}
\end{equation}
If one substitutes the series (\ref{v4.34}) into the above equation, and then utilizes the series (\ref{v2.9}) and (\ref{v2.10}) for $\gamma(x)$
and $\beta(x)$, one finds after a little algebra that
{\allowdisplaybreaks
\begin{gather}
T_{1,1}  =  1, \; T_{2,2} = \left(1 + \beta_0\right) / 2 \;
\stackrel{n_f = 5}{\longrightarrow}
\frac{35}{24},
\label{t11}
\\
T_{2,1}  =  \left(1 + \beta_0\right) T_{1,0} + \gamma_1 - 2 \;
\stackrel{n_f = 5}{\longrightarrow}
5.4208
\\
T_{3,3}  =  \left(1 + \beta_0\right) \left(1 + 2 \beta_0\right) / 6 \;
\stackrel{n_f = 5}{\longrightarrow}
\frac{1015}{432},
\\
\begin{split}
T_{3,2}  &=  \left(1 + \beta_0\right)\left(1 + 2 \beta_0\right) T_{1,0} / 2 + \beta_0 \left(\gamma_1 - 3\right)
+ \gamma_1 + \frac{\beta_1}{2} - 2
\\*
&\stackrel{n_f = 5}{\longrightarrow}
13.1053,
\end{split}
\\
\begin{split}
T_{3,1}  &=  \left(1 + 2 \beta_0\right) T_{2,0} + \left(\beta_1 + \gamma_1 - 2 \beta_0 - 2\right) T_{1,0}
+ \gamma_2 - 4 \gamma_1 + 4
\\*
& \stackrel{n_f = 5}{\longrightarrow}
 42.3366
 \end{split}
\label{v4.37}
\end{gather}
}

We then can employ the three-loop series
\begin{equation}
T[x, L] = \sum_{j=1}^3 \sum_{k=0}^j T_{j,k} x^j L^k
\label{v4.38}
\end{equation}
with series coefficients given in Eqs.\ (\ref{v4.35}) and (\ref{t11})--(\ref{v4.37}) to compare the pole mass obtained from the
extracted value $m_b (10\,{\rm GeV})$ to that from the correspondingly RG-evolved value $m_b \left(m_b\right)$.  To be consistent with having
three subleading orders in $x$ in series (\ref{v4.34}), we utilize Eqs.\ (\ref{v2.9}) and (\ref{v2.10}) to four-loop order to evolve $x(\mu)$
and $m_b (\mu)$ [$\beta_3 = 18.8522$ \cite{beta2}, $\gamma_3 = 11.0343$ \cite{gamma2}] from the same
$x(10 \; {\rm GeV}) = 0.056732$
reference couplant value [as evolved from an assumed $\alpha_s\left(M_Z\right)=0.11800$] used throughout.
Using the $\mu = 10 \; {\rm GeV}$, $N=2$ value $m_b^{(2)} (10\,{\rm GeV}) = 3.651 \; {\rm GeV}$ of Table \ref{vtab2} as a springboard
value, and its corresponding value $m_b \left(m_b\right) = 4.19\,{\rm GeV}$ (Table \ref{vtab3}), we find somewhat different pole
masses for different choices of $\mu$:
\begin{gather}
\mu = 10 \; {\rm GeV}, \; \; \; \; M_b^p = 4.82 \; {\rm GeV},
\label{v4.39}
\\
\mu = 4.19 \; {\rm GeV}, \; \; \; \; M_b^p = 4.94 \; {\rm GeV},
\label{v4.40}
\end{gather}
indicative of $120\, {\rm MeV}$ residual scale uncertainty.  We emphasize that this uncertainty arises entirely from the truncation
of the series (\ref{v4.34}), and is independent of scale uncertainties in the extraction of $m_b (\mu)$.  Had we used the corresponding
$N = 2$, $\mu = 10 \; {\rm GeV}$ values $m_b (10\,{\rm GeV}) = 3.665 \; {\rm GeV}$, $m_b \left(m_b\right) = 4.21 \; {\rm GeV}$
obtained in Tables 4 and 5 via optimal RG improvement, the corresponding pole masses still exhibit virtually the same residual
scale uncertainty:
\begin{gather}
\mu = 10 \; {\rm GeV}, \; \; \; \; M_b^p = 4.84 \; {\rm GeV},
\label{v4.41}\
\\
\mu = 4.21 \; {\rm GeV}, \; \; \; \; M_b^p = 4.95 \; {\rm GeV}.
\label{v4.42}
\end{gather}
Consequently, there appears to be a surprisingly large $110\mbox{--}120 \, {\rm MeV}$ theoretical uncertainty implicit in
the determination of the pole mass from the $\overline{{\rm MS}}$ mass $m_b (\mu)$, an uncertainty devolving ultimately from
truncation of the series (\ref{v4.34}) after its $\mathcal{O}\left(x^3\right)$ terms. In the section which follows, we will
optimally RG-improve the series (\ref{v4.34}) to reduce this residual scale uncertainty by a factor of 15.

\section{Optimal RG Improvement of the Pole Mass}
\label{resum_pole_sec}
Optimal RG-improvement of the series $T [x, L]$ follows along the same lines as described in Section 3 for the
series $S_N [x, L]$.  We express the series (\ref{v4.34}) in the form
\begin{equation}
T[x, L] = \sum_{n=0}^\infty T_n (xL) x^n
\label{v5.43}
\end{equation}
where
\begin{equation}
T_n (u) = \sum_{k=n}^\infty T_{k, k-n} u^{k-n}
\label{v5.44}
\end{equation}
with $T_0 (xL)$ encompassing the LL summation (all values of $T_{k,k}$), $T_1 (xL)$ encompassing the NLL summation,
{\it etc}.  Since $T_{3,0}$ is known [Eq.\ (\ref{v4.35})], the functions $T_0 (xL)$, $T_1 (xL)$, $T_2 (xL)$ and the ${\rm N^3 LL}$
summation $T_3 (xL)$ within Eq.\ (\ref{v5.43}) are all RG accessible.  If we substitute Eq.\ (\ref{v5.43}) into the RG-equation (\ref{v4.36}), we
find the following set of sequentially solvable first order differential equations for $T_n (u)$, $n = \{0, 1, 2, 3 \}$,
with initial conditions $T_n (0) = T_{n,0}$ [Eq.~(\ref{v5.44})] explicitly listed in Eq.\ (\ref{v4.35}):
{\allowdisplaybreaks
\begin{gather}
\left[ \left(1 - \beta_0 u\right) \frac{d}{du} - 1 \right] T_0 (u) = 0,
\label{v5.45}
\\
\left[ \left(1 - \beta_0 u\right) \frac{d}{du} - \left(1 + \beta_0\right) \right] T_1 (u)
= \left[ \left(\beta_1 u - 2\right) \frac{d}{du} + \gamma_1 \right] T_0 (u),
\label{v5.46}
\\
\begin{split}
0=&-\left[ \left(1 - \beta_0 u\right) \frac{d}{du} - \left(1 + 2 \beta_0\right) \right] T_2 (u)
+ \left[ \left(\beta_1 u - 2\right) \frac{d}{du} + \left(\gamma_1 + \beta_1\right) \right] T_1 (u)
\\*
&+ \left[ \left(\beta_2 u - 2\gamma_1\right) \frac{d}{du} + \gamma_2 \right] T_0 (u),
\end{split}
\label{v5.47}
\\
\begin{split}
0=&-\left[ \left(1 - \beta_0 u\right) \frac{d}{du} - \left(1 + 3 \beta_0 \right) \right] T_3 (u)
+ \left[ \left(\beta_1 u - 2 \right) \frac{d}{du} + \left(\gamma_1 + 2\beta_1\right) \right] T_2 (u)
\\*
&+ \left[ \left(\beta_2 u - 2 \gamma_1\right) \frac{d}{du} + \left(\gamma_2 + \beta_2 \right) \right] T_1 (u)
+ \left[ \left(\beta_3 u - 2 \gamma_2 \right) \frac{d}{du} + \gamma_3 \right] T_0 (u).
\end{split}
\label{v5.48}
\end{gather}
}
Given $u = xL$ and the definition $w \equiv 1 - \beta_0 xL$, the solutions to the above four equations are
{\allowdisplaybreaks
\begin{gather}
T_0 (xL) = w^{-A},
\label{v5.49}
\\
T_1 (xL) = B w^{-A} + w^{-A-1} \left[ T_{1,0} - B + C \log (w) \right],
\label{v5.50}
\\
\begin{split}
T_2 (xL)  = & \frac{D}{2} w^{-A} + w^{-A-1} \left[ E-F + F \log (w)\right]
\\*
& +  w^{-A-2} \left[ T_{2,0} - \frac{D}{2} - E + F + G \log (w) + \frac{H}{2} \log^2 (w) \right],
\end{split}
\label{v5.51}
\\
\begin{split}
T_3 (xL)  = & \frac{K}{3}w^{-A} + w^{-A-1} \left( \frac{M}{2} - \frac{N}{4} + \frac{N}{2} \log (w) \right)
\\*
& +  w^{-A-2} \left[ P - Q + 2R + (Q-2R)\log (w) + R \log^2 (w) \right]
\\*
& +  w^{-A-3} \Biggl\{ -\frac{K}{3} - \frac{M}{2} + \frac{N}{4} - P + Q - 2R + T_{3,0} + U \log (w) \Biggr.
\\*
& \qquad\qquad+  \Biggl. \frac{V}{2} \log^2 (w) + \frac{Y}{3} \log^3 (w) \Biggr\},
\end{split}
\label{v5.52}
\end{gather}
}
where parameters $A$ -- $Y$ are now given by
{\allowdisplaybreaks
\begin{gather}
A = 1/\beta_0, \; \; B = \left(A\beta_1 - \gamma_1\right)/\beta_0, \; \; C = \left(2\beta_0 - \beta_1\right) A / \beta_0,
\label{v5.53}
\\
D = \left[ \beta_1 AB + \beta_2 A - \left(\beta_1 + \gamma_1\right) B - \gamma_2 \right] / \beta_0,
\label{v5.54}
\\
\begin{split}
\beta_0 E  = & \left(2\beta_0 - \beta_1\right) AB + \left[ (1+A)\left(T_{1,0} - B\right) - C \right] \beta_1
\\* & +   \left(2\beta_0 \gamma_1 - \beta_2\right) A + \left(B-T_{1,0}\right)\left(\beta_1 + \gamma_1\right)
,
\end{split}
\label{v5.55}
\\
F = \left(A\beta_1 - \gamma_1\right) C/\beta_0,
\label{v5.56}
\\
G = \left[ (1+A)\left(T_{1,0} - B\right) - C\right]\left(2\beta_0 - \beta_1\right) / \beta_0,
\label{v5.57}
\\
H = \left(2\beta_0 - \beta_1\right)(1+A)C/\beta_0,
\label{v5.58}
\\
\beta_0K  =  \left[ A\left(\beta_3 + \beta_2 B + \beta_1 D / 2\right) - B\left(\gamma_2 + \beta_2\right) - \gamma_3
 +   \left(\gamma_1 + 2\beta_1\right)D/2 \right],
\label{v5.59}
\\
\begin{split}
\beta_0 M  = &  \left[ \left(2\beta_0 \gamma_2 - \beta_3\right) + B \left(2\beta_0 \gamma_1 - \beta_2\right)
+ D \left(2\beta_0 - \beta_1\right)/2\right]A
\\*
& +  \left[ \gamma_2 + \beta_2 - (A+1)\beta_2 \right] \left(B - T_{1,0}\right) - C\beta_2
\\*
& +  \left(\gamma_1 + 2 \beta_1\right) (F-E) + \left[ E(1+A) - F(2+A)\right] \beta_1 ,
\end{split}
\label{v5.60}
\\
N = \left[ C \left(A \beta_2 - \gamma_2\right) + F \left[ (A-1)\beta_1 - \gamma_1 \right] \right] / \beta_0,
\label{v5.61}
\\
\begin{split}
\beta_0P  = &  \left(2\beta_0 \gamma_1 - \beta_2\right) \left[ (1 + A)\left(T_{1,0} - B\right) - C \right]
\\*
 &+  \left(2\beta_0 - \beta_1\right) \left[ (1+A) E-(2+A)F \right]
 \\*
& -  \left(\gamma_1 + 2 \beta_1\right) \left(T_{2,0} - {D}/{2} - E + F\right)
\\*
 &-   \beta_1 \left[ G - (2+A) \right]\left(T_{2,0} - {D}/{2} - E + F\right),
\end{split}
\label{v5.62}
\\
\begin{split}
\beta_0Q  = &  \left[ \left(2\beta_0 \gamma_1 - \beta_2\right) C + \left(2\beta_0 - \beta_1\right) F \right] (1+A)
\\*
& -  \left(\gamma_1 + 2\beta_1\right) G - \left[ H - (2+A) G\right] \beta_1 ,
\end{split}
\label{v5.63}
\\
R = \left(\beta_1 A - \gamma_1\right) H / 2 \beta_0,
\label{v5.64}
\\
U = \left(2\beta_0 - \beta_1\right) \left[ (2+A)\left(T_{2,0} - D/2 - E + F\right) - G\right] / \beta_0,
\label{v5.65}
\\
V = \left(2\beta_0 - \beta_1 \right) \left[ (2+A) G - H \right] / \beta_0,
\label{v5.24}
\\
Y = \left(2\beta_0 - \beta_1\right)(2+A) H / 2\beta_0.
\label{v5.67}
\end{gather}
}

We now can compare the pole mass obtained via Eq.\ (\ref{v4.33}) from the first four terms of the series (\ref{v5.43}), which include all RG-accessible coefficients,
\begin{equation}
T[x,L]  =  T_0 (xL) + x T_1 (xL)
 +  x^2 T_2 (xL) + x^3 T_3 (xL),
\label{v5.68}
\end{equation}
to the pole mass obtained via the three-loop series (\ref{v4.38}).  If we utilize Table \ref{vtab5}'s $N=2$, $\mu = 10 \; {\rm GeV}$ extracted value $m_b (10 \; {\rm GeV}) = 3.665 \; {\rm GeV}$ and incorporate Eq.\ (\ref{v5.68}) into Eq.\ (\ref{v4.33}), we find that
\begin{gather}
\mu = 10 \; {\rm GeV}, \; \; \; M_b^p = 4.959 \; {\rm GeV}
\label{v5.69}
\\
\mu = 4.207 \; {\rm GeV}, \; \; \; M_b^p = 4.951 \; {\rm GeV}
\label{v5.70}
\end{gather}
This $\mathcal{O} (8 \; {\rm MeV})$ uncertainty is
a remarkable improvement over
 the $\mathcal{O}(110 \; {\rm MeV})$ uncertainty
[Eqs.\ (\ref{v4.41}) and (\ref{v4.42})]
following from the same input assumptions, but with $T[x,L]$ given by the three loop series (\ref{v4.38}).  Thus, the
RG-improved series (\ref{v5.68}) removes virtually all the residual scale dependence in the relation between the pole and
$\overline{{\rm MS}}$ mass. This is corroborated in Fig.\ \ref{fig4}, in which the pole masses obtained via the three loop series
(\ref{v4.38}) and the RG-improved series (\ref{v5.68}) are compared directly, given identical
$m_b (10\,{\rm GeV}) = 3.6636 \; {\rm GeV}$ anchoring values for the $\overline{{\rm MS}}$ mass [the $\mu = 10 \; {\rm GeV}$ value for
$\overline{m_b} \left(m_b\right)$ in Table \ref{vtab4}].  The pole mass for the RG-improved case exhibits very little dependence
on $\mu$.  Thus optimal RG-improvement of the perturbative series (\ref{v4.34}) is seen to remove virtually all of the substantial
($\mathcal{O}(110\mbox{--}120 \; {\rm MeV})$) residual uncertainty characterizing the relationship between the $\overline{{\rm MS}}$
and the pole $b$-quark mass.

\begin{figure}[hbt]
\centering
\includegraphics[scale=0.55]{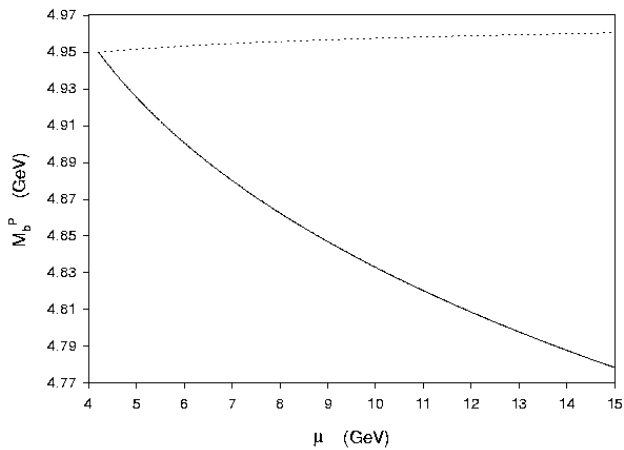}
\caption{
 Renormalization scale dependence of the resummed expression of the pole mass (broken curve) compared with the
 unsummed expression (solid curve).  The $\overline{\rm MS}$ mass $m_b(\mu)$ used as input to this comparison
 is based on RG evolution from the reference value $m_b(10\,{\rm GeV})=3.66\,{\rm GeV}$ as outlined in the text.
 }
\label{fig4}
\end{figure}

\section{Optimal RG Improvement of $M_b^{1S}/ M_b^p$}
\label{1S_sec}
The $b$-quark $1S$ mass $\left(M_b^{1S}\right)$, defined to be half the perturbative mass of a (theoretical) $3S_1 \; b\bar{b}$
meson, has been determined via relations between masses and widths of $\Upsilon$ mesons to moments of the $b$-quark
vector current correlation function \cite{ligeti}.  This $1S$ mass is related to the pole mass $M_b^p$ (which is much more
sensitive to $\Lambda_{QCD}$) via the perturbative relationship \cite{1S,hoang},
\begin{equation}
M_b^{1S} = M_b^p \left[ 1 - \frac{2\pi^2 x^2 (\mu)}{9} W(\mu) \right],
\label{v6.71}
\end{equation}
for which the series $W(\mu)$ is known in full to two subleading orders:
\begin{gather}
W(\mu)  = W\left[ x(\mu), \; \ell(\mu)\right]\quad , \quad  \ell(\mu) = \log \left[ \frac{3\mu}{4\pi x(\mu) M_b^p}
\right],
\\
W[x,\ell]  =  1 + \sum_{n=1}^\infty \sum_{m=0}^n \tau_{n,m} x^n \ell^m
\label{v6.72}
\end{gather}
where
{\allowdisplaybreaks
\begin{gather}
\tau_{1,0}  =  \frac{97}{6} - \frac{11 n_f}{9}
\stackrel{n_f = 5}{\longrightarrow}
\frac{181}{18}
\\
\tau_{2,0}  =  337.95 - 40.965 n_f + 1.1629 n_f^2
\stackrel{n_f = 5}{\longrightarrow}
 162.19
 \\
\tau_{1,1}  =  4 \beta_0
\stackrel{n_f = 5}{\longrightarrow}
 \frac{23}{3}
\label{tau11}
\\
\tau_{2,1}  =  6\beta_0 \tau_{1,0} - 8\beta_0^2 + 4\beta_1
\stackrel{n_f = 5}{\longrightarrow}
 \frac{1151}{12}
\\
\tau_{2,2}  =  12 \beta_0^2
\stackrel{n_f = 5}{\longrightarrow}
 \frac{529}{12}.
\label{v6.73}
\end{gather}
}
In principle, $M_b^{1S}$ and $M_b^p$ are both RG-invariant
entities independent of the renormalization scale $\mu$.  Consequently, one can obtain the following RG-equation for $W[x,\ell]$ by
requiring that $\frac{d}{d\mu^2} \left[ x^2 (\mu) W(\mu)\right]=0$: \begin{equation} \left[ 2\beta(x) +
x\beta(x)\frac{\partial}{\partial x} + \left[ \frac{x}{2} - \beta(x) \right] \frac{\partial}{\partial \ell} \right]W[x,\ell] = 0,
\label{v6.74} \end{equation}
with $\beta(x)$ given by Eq.\ (\ref{v2.10}).  If one substitutes the series (\ref{v6.72}) into the RG-equation (\ref{v6.74}), one easily corroborates the results
(\ref{tau11})--(\ref{v6.73}).

Optimal RG improvement of the series $W[x,\ell]$ is obtained by expressing the series in the form
\begin{equation}
W[x,\ell] = \sum_{n=0}^\infty x^n W_n (x\ell)
\label{v6.75}
\end{equation}
where
\begin{equation}
W_n (x\ell) = \sum_{k=n}^\infty \tau_{k, k-n} (x\ell)^{k-n} .
\label{v6.76}
\end{equation}
As before, the series $W_0 (x\ell)$ is inclusive of all LL coefficients $\tau_{k,k}$ in the series (\ref{v6.72}).  Similarly,
$W_1(x\ell)$ includes all NLL coefficients $\tau_{k, k-1}$, and $W_2 (x\ell)$ includes all
NNLL coefficients $\tau_{k, k-2}$.  Upon substituting Eq.\ (\ref{v6.75}) into the RG equation (\ref{v6.74}), we obtain
successive first order differential equations for the $n = \{0,1,2\}$ cases of series (\ref{v6.76}):
{\allowdisplaybreaks
\begin{gather}
\left[ \left(1 - 2\beta_0 u\right) \frac{d}{du} - 4\beta_0 \right] W_0 (u) = 0,
\label{v6.77}
\\
\left[ \left(1 - 2\beta_0 u\right) \frac{d}{du} - 6\beta_0\right] W_1 (u)
= \left[2\left(\beta_1 u - \beta_0\right) \frac{d}{du} + 4\beta_1 \right] W_0 (u),
\label{v6.78}
\\
\begin{split}
\left[ \left(1 - 2 \beta_0\right) \frac{d}{du} - 8\beta_0 \right] W_2 (u)
 = & \left[ 2\left(\beta_1 u - \beta_0\right) + 6\beta_1 \right] W_1 (u)
\\*
 &+  \left[ 2\left(\beta_2 u - \beta_1\right) +4\beta_2 \right]W_0 (u).
\end{split}
\label{v6.79}
\end{gather}
}
Given the initial conditions $W_0 (0) = \tau_{0,0} = 1$, $W_1 (0) = \tau_{1,0}$ and $W_2 (0) = \tau_{2,0}$ evident from the definition (\ref{v6.76}) of $W_n (u)$, we obtain the following solutions to Eqs.\ (\ref{v6.77}), (\ref{v6.78}) and (\ref{v6.79}):
\begin{gather}
W_0 (x\ell) = \left(1 - 2\beta_0 x\ell\right)^{-2}
\label{v6.80}
\\
W_1 (x\ell) = \left(1 - 2\beta_0 x\ell\right)^{-3} \left[ \tau_{1,0}
+ 4 \left( \beta_0 - \frac{\beta_1}{2\beta_0} \right) \log \left( 1 - 2\beta_0 x\ell\right) \right],
\label{v6.81}
\\
\begin{split}
W_2 (x\ell)  = & \mathcal{U} \left(1 - 2\beta_0 x\ell\right)^{-3}
\\
& +   \frac{\left[ \tau_{2,0} - \mathcal{U}
+ \mathcal{VX} \log \left(1 - 2\beta_0 x\ell\right)
 +   3\mathcal{V}^2 \log^2 \left(1 - 2\beta_0 x\ell\right) \right]}{\left(1 + 2\beta_0 x\ell\right)^{4}},
\end{split}
\label{v6.82}
\end{gather}
where
\begin{gather}
\mathcal{U} = 2\left[ \frac{\beta_1^2}{\beta_0^2} - \frac{\beta_2}{\beta_0} \right],
\label{v6.83}
\\
\mathcal{V} = \left(\beta_1 - 2\beta_0^2\right) / \beta_0,
\label{v6.84}
\\
\mathcal{X} = -3\tau_{1,0} + 4 \beta_0 - 2\beta_1 / \beta_0,
\label{v6.85}
\end{gather}
with $\tau_{1,0}$ and $\tau_{2,0}$ as given in Eq.\ (\ref{v6.73}).

Let us first consider any residual scale dependence in the relation (\ref{v6.71}) arising from truncation of the
series (\ref{v6.72}) after the known coefficients (\ref{v6.73}).  It has been argued  in Ref.\ \cite{hoangc} that the relation
(\ref{v6.71}) is to be utilized at soft momentum scales ($1.5 \, {\rm GeV} < \mu < 3.5 \, {\rm GeV}$), because  the
$1S$ mass is defined purely from nonrelativistic dynamics.  Moreover, a soft scale necessarily follows from
the non-perturbative nature of the logarithm $\ell$;  if $\mu$ is hard, then $x$ is small and logarithms are large.  Given Eq.\
(\ref{v5.69})'s $b$-quark pole mass of $4.96 \; {\rm GeV}$, for example, we find for the truncated series that the $1S$ $b$-quark mass
varies between $4.58$ and $4.73 \; {\rm GeV}$ as $\mu$ increases from $1.5$ and $3.5 \; {\rm GeV}$, a $150 \; {\rm MeV}$ residual scale
uncertainty.   This range is substantially diminished if we utilize the RG-improved series
\begin{equation}
W[x,\ell] = W_0 (x\ell) + x W_1 (x\ell) + x^2 W_2 (x\ell),
\end{equation}
as determined by Eqs.\ (\ref{v6.80}), (\ref{v6.81}) and (\ref{v6.82}), for the same region of $\mu$.  For the RG
improved series, we find that the $1S$ $b$-quark mass corresponding to a $4.96 \; {\rm GeV}$ pole mass varies
between $4.66\, {\rm GeV}$ and $4.61 \; {\rm GeV}$ as $\mu$ increases from $1.5\,{\rm GeV}$ to $3.5 \; {\rm GeV}$.  Thus, optimal RG
improvement reduces the residual scale uncertainty from $150 \; {\rm MeV}$ to $50 \; {\rm MeV}$.  These results are presented
graphically in Fig.\ \ref{fig5}.

\begin{figure}[hbt]
\centering
\includegraphics[scale=0.55]{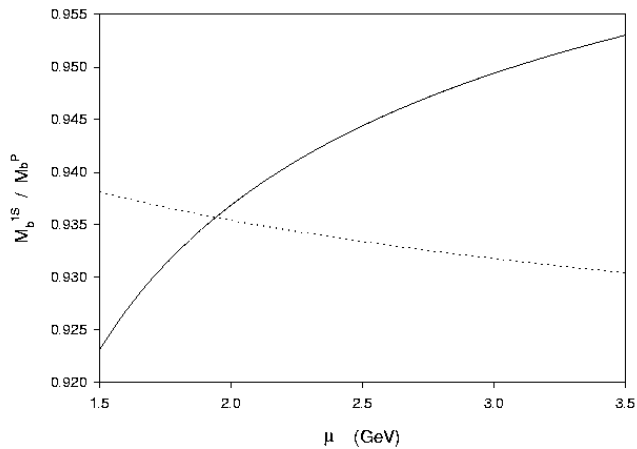}
\caption{
 Renormalization scale dependence of the resummed expression of the ratio $M_b^{1S}/M_b^p$ (broken curve) compared with the
 unsummed expression (solid curve).
  The range considered for  $\mu$ is the ``soft'' region advocated in \protect\cite{hoangc}, and
   $M_b^p=4.96\,{\rm GeV}$ is used as an input value.
 }
\label{fig5}
\end{figure}

It is, of course, more realistic to extract a $b$-quark pole mass from a phenomenological determination of the
$1S$ $b$-quark mass.  Using the central value $4.71 \; {\rm GeV}$ for the $1S$ mass \cite{hoangc}, one can invert the relation
(\ref{v6.71}) numerically for both the truncated and the RG-improved versions of the series (\ref{v6.72}).  The
results of this inversion are displayed in Fig.\ \ref{fig6}, and are indicative of a pole mass somewhat above $5 \; {\rm GeV}$.
Once again, however, an $\mathcal{O} (150 \; {\rm MeV})$ theoretical scale uncertainty for the truncated case is
reduced to a $40 \; {\rm MeV}$ scale uncertainty using the RG improved series.  Note that the crossing point in both
figures is the soft-$\mu$ point at which the logarithm $\ell$ is equal to zero.  It is evident from the initial conditions for
$W_0$, $W_1$ and $W_2$ that the RG summed series and the truncated series are equivalent at this point.

\begin{figure}[hbt]
\centering
\includegraphics[scale=0.55]{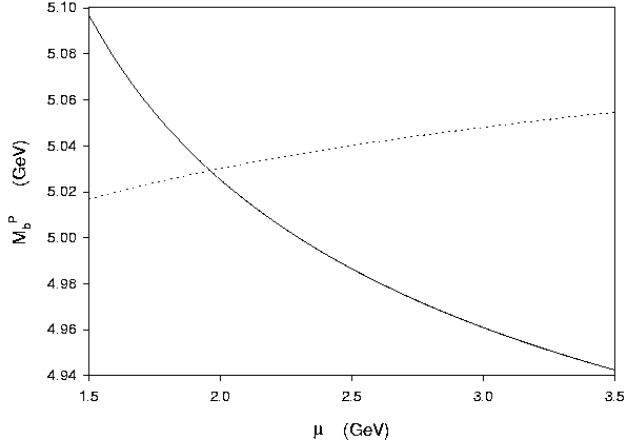}
\caption{
 Renormalization scale dependence of the resummed expression for  $M_b^p$ (broken curve) compared with the
 unsummed expression (solid curve).
  The input value $M_b^{1S}=4.71\,{\rm GeV}$ and the range considered for  $\mu$ follows from
Ref.\ \protect\cite{hoangc}.
 }
\label{fig6}
\end{figure}

\section{Conclusions}
As demonstrated in Section \ref{scale_sec}, the procedure for extracting the $\overline{\rm MS}$ $b$-quark mass from empirical moments
$M_N$ of $R(s)$ necessarily exhibits theoretical dependencies on the choice of renormalization scale ($\mu$), the choice of moment
($N$), and the effect of higher-order perturbative contributions. Omitting coupling-constant and experimental uncertainties, we have
shown that an analysis based upon the Ref.\ \cite{KS} choices $\mu = 10\,{\rm  GeV}$ and $N = 2$ leads to the extraction of an
$\overline{\rm MS}$ $b$-quark mass
\begin{equation}
     m_b\left(m_b\right) = 4.193 \,{\rm GeV}\pm  15\,{\rm  MeV}\pm   8.5\, {\rm MeV} \pm   20\, 
     {\rm MeV}.
     \label{final_uncert}
\end{equation}
The first theoretical uncertainty is associated with a $\pm 5 \,{\rm GeV}$ variation of the renormalization
scale $\mu$ \cite{KS},  the
second  reflects the moment dependence in letting the choice of moment $N$ vary from 1 to 4, and the third is an estimate of
higher-order
perturbative contributions.

 In Section \ref{resum_sec}, the perturbative series (\ref{v2.5}) from which this prediction is obtained
is optimally
RG-improved via the all-orders summation of that series' leading, next-to-leading, and
next-to-next-to-leading logarithms, {\it i.e.,} the summation of all RG-accessible logarithms in the perturbative
series.
In addition, an estimated value for $T^{(N)}_{3,0}$ based on the approximate power-law growth of the RG-undetermined perturbative
coefficients allows an all-orders summation to a further subleading-logarithm order.
If input assumptions leading to Eq.\ (\ref{final_uncert}) are otherwise unchanged, the optimally RG-improved
$\overline{\rm MS}$ mass is then found to be
\begin{equation}
     m_b\left(m_b\right) = 4.207 \,{\rm GeV} \pm 0.3 \,{\rm MeV} \pm 4.2\,{\rm MeV} \pm 7\,{\rm MeV}.
\end{equation}
As before, the renormalization-scale and moment-dependence uncertainties displayed above are associated respectively with varying $\mu$
by $\pm 5\, {\rm GeV}$, with varying $N$ from 1 to 4, and with the effect of higher-order perturbative contributions as estimated via
(\ref{T30_eq}).

     The reduction in these latter uncertainties associated with the choice of moment $N$ and the (estimated) higher-order perturbative
contributions is an unanticipated but welcome feature of the RG-summation developed in Section \ref{resum_sec}.  Indeed,
{\em prior to
such improvement}, the uncertainty ($\sigma_N$) devolving from varying $N$ is seen in Table \ref{vtab3} to increase quite
drastically with $\mu$ ($\sigma_N = 35\, {\rm MeV}$ at $\mu = 15\, {\rm GeV}$). After RG-improvement, however, this
$N$-uncertainty is reduced to $4\,{\rm  MeV}$ levels regardless of the choice for $\mu$ (Table \ref{vtab5}).
Since the resummation analysis reduces the theoretical errors to levels negligible compared with those of experimental inputs
($M^{exp}_N$ \cite{KS} and $\alpha_s\left(M_Z\right)=0.119\pm 0.002$ \cite{pdg}), the benchmark $m_b\left(m_b\right)$ determination
\cite{KS} is modified to \begin{equation}
m_b\left(m_b\right)=4.207\,{\rm GeV}\pm 40\,{\rm MeV}\quad .
\end{equation}
With the reduced theoretical uncertainties and with use of the range  $\alpha_s\left(M_Z\right)=0.119\pm 0.002$ \cite{pdg}, the
dominant source of uncertainty now arises from the experimental moments $M^{exp}_N$.

     Renormalization scale dependence inherent in relations between the $b$-quark pole mass and
corresponding $b$-quark $\overline{\rm MS}$ and $1S$ masses is shown in Sections \ref{resum_pole_sec} and
\ref{1S_sec} to be similarly reduced via
optimal RG-improvement of the perturbative series characterizing such relations. Comparison of
``RG-unimproved" extractions of the pole mass [Eqs.\ (\ref{v4.41}) and (\ref{v4.42})] to RG-improved extractions
[Eqs.\ (\ref{v5.69}) and (\ref{v5.70})] for which all input information is otherwise equivalent indicates a reduction
from $110\, {\rm MeV}$ to $8 \, {\rm MeV}$ in the variation of the pole mass with renormalization scale as that scale
varies from $4.2 \, {\rm GeV}$ to $10 \, {\rm GeV}$. Moreover, for the improved case, the central-value pole mass is
found to be near the high end of the range for the ``unimproved" pole mass. A similar elevation with
RG-improvement characterizes the pole mass extracted from the $1S$ mass, as discussed in Section
\ref{1S_sec}. In Fig.\ \ref{fig6}, the summation of RG-accessible logarithms is shown to lead to a less
scale-dependent and somewhat larger pole mass extracted from an assumed $4.71 \, {\rm GeV}$ $b$-quark
$1S$-mass than would occur in the absence of such RG-improvement.
In particular, there is a reduction from $140\,{\rm MeV}$ to $40\,{\rm MeV}$ in the variation of the pole mass as
the renormalization scale $\mu$ varies between $1.5\,{\rm GeV}$ and $3.5\,{\rm GeV}$.  To our knowledge, such
renormalization-scale theoretical uncertainties have not previously been considered in detail.

The important point common to all of the cases considered above, however, is that (often-ignored)
theoretical uncertainties necessarily follow from the residual renormalization-scale dependence
characterizing the truncation of  phenomenological  perturbative series, and that such uncertainties
may be substantially reduced, if not eliminated, by improving such series to include summation of
all higher order RG-accessible contributions.  Such resummation techniques should thus prove to  be of
increasing value
as phenomenological and experimental inputs into $b$-mass determinations become more precise.

\section{Acknowledgments}
Research funding from the Natural Science and Engineering Research Council
of Canada (NSERC) is gratefully acknowledged. Ailin Zhang is partly
supported by National Natural Science Foundation of China.

\end{document}